\documentclass[twocolumn,showpacs,preprintnumbers,amsmath,amssymb]{revtex4}
\usepackage{graphicx}% Include figure files
\usepackage{dcolumn}% Align table columns on decimal point
\usepackage{bm}% bold math

\usepackage{amsfonts}
\usepackage{amsthm}
\usepackage{amssymb}
\usepackage{esint}
\usepackage{tipx}
\usepackage[mathscr]{euscript}
\usepackage{t1enc}
\usepackage{amsmath}
\usepackage{amscd}
\usepackage{psfrag}
\usepackage{graphics}

\begin{document}

%\preprint{APS/123-QED}

\title{Functional integral for non-Lagrangian systems}% Force line breaks with \\

\author{Denis Kochan}
\email{kochan@fmph.uniba.sk}
\affiliation{%
\small{Department of Theoretical~Physics, \\
Comenius University, Bratislava, Slovakia \\
$\&$ \\
Theory Group, CERN,\\
Gen\`{e}ve, Switzerland}}

%\date{\today}

\begin{abstract}
\noindent A novel functional integral formulation of quantum
mechanics for non-Lagrangian systems is presented. The new
approach, which we call ``stringy quantization,'' is based solely
on classical equations of motion and is free of any ambiguity
arising from Lagrangian and/or Hamiltonian formulation of the
theory. The functionality of the proposed method is demonstrated
on several examples. Special attention is paid to the stringy
quantization of systems with a general $A$-power friction force
$-\kappa\dot{q}^A$. Results for $A = 1$ are compared with those
obtained in the approaches by Caldirola-Kanai, Bateman and Kostin.
Relations to the Caldeira-Leggett model and to the Feynman-Vernon
approach are discussed as well.

\vspace{4mm}
\centerline{Dedicated to my father on the occasion of his 60th birthdays.}
\end{abstract}

\maketitle

\section{Introduction}\label{1}

Quantization is a phenomenon that changes our bright classical
perspective into a bit uncertain and at first sight rather
nonintuitive picture. This picture, however, is more rigorous than
the classical one, possesses many fascinating features and has
produced a lot of successful predictions.

The subtle problem of transition from classical to quantal
attracts attention from the early days of quantum mechanics. Over
the years various techniques and methods for solving this puzzle
have been invented. Our aim is not to trace back the complete
(hi)story of the milestone ideas in this field (for the review we
refer to \cite{englis}). What we want to do is to give a concise
exposition of the method we have developed. However, since we have
generalized the original Feynman's path integral approach, we will
recapitulate this approach shortly in section \ref{2}.

The main goal of our paper is to obtain a functional integral
formula for the quantum propagator which would not refer to
Lagrangian and/or Hamiltonian function. Quantum propagator is the
probability amplitude $\mathbf{A}(q_1,t_1\,|\,q_0,t_0)$ for the
transition of the system from the initial configuration
$(q_0,t_0)$ to the final configuration $(q_1,t_1)$. We will derive
a closed expression for this quantity starting from the given set
of classical dynamical equations of motion.

The proposed method uses functional integration in the extended
phase space, but instead of integration over path histories we
introduce integration over stringy surfaces. This crucial element
of our approach is explained in full detail in sections \ref{3}
and \ref{4} and in appendices \ref{A} and \ref{B}. We also make
sure that whenever the system under consideration becomes
Lagrangian, the stringy description reduces to the standard one
with the Feynman path integral.

In sections \ref{4.1} and \ref{4.2} some simple examples are
scrutinized. It is well known that there are classical dynamical
systems that cannot be described within the traditional Lagrangian
or Hamiltonian framework. The stringy approach enables us to
quantize them straightforwardly. In section \ref{4.1} the
quantization of a weakly non-Lagrangian system is performed. The
transition amplitude is computed for a particle in a conservative
field, whose motion is damped by a general $A$-power friction
force $F=-\kappa\dot{q}^{A}$. The stringy results for $A=1$ are
compared with the results obtained in the approaches by
Caldirola-Kanai \cite{CK}, Bateman \cite{bateman} and Kostin
\cite{kostin}.

One can argue against the stringy quantization of dissipative
systems that it is unable to describe decoherence phenomena. This
is true, but the same objection can be raised against the
generally accepted heuristic approaches by the authors cited
above, as well as those by Dekker \cite{dekker2}, Razavy
\cite{razavy2}, Geicke \cite{geicke} and others, simply because
their kinematical and dynamical prerequisites are different from
the prerequisites of the particle-plus-environment quantum models.
However, a possible argument for the stringy quantization is that
the particle-plus-environment models are not able to handle
satisfactorily the case with the friction force $-\kappa\dot{q}^A$
for the general power $A$. Moreover, they describe a rather
different phenomenon, namely the quantum Brownian motion for which
the total force equals $-\kappa\dot{q}+\mbox{\emph{stochastic
term}}$.

Section \ref{4.2} contains the analysis of a curious
two-dimensional Douglas system \cite{douglas} which is strongly
non-Lagrangian, i.e. not derivable from any sort of Lagrangian or
Hamiltonian. This causes a fundamental problem for all
conventional quantization methods, but can be dealt with in a
rather transparent way in the stringy approach.

In section \ref{5} conclusion, discussion and outlook are
collected. The section also includes some comments on the relation
between the stringy quantization and the Caldeira-Leggett model of
quantum Brownian motion \cite{particle+environment}, as well as
the influence functional technique by Feynman and Vernon
\cite{feynman-vernon}.

Some rather technical material is left to appendices. Appendix
\ref{A} is devoted to the stringy variational principle, which
plays an important role in the motivation of our approach. In
appendix \ref{B} computational details concerning the surface
functional integral for quantum friction force systems are
presented.

Historically the first attempt at quantization based on the
dynamical equations of motion belongs to Feynman, see
\cite{dyson}. A similar problem was considered by Wigner, Yang
$\&$ Feldman, Nelson, Okubo and others, see \cite{w-yf-o}. Among
the recent investigations in this field let us mention the work of
Lyakhovich $\&$ Sharapov \cite{sharapov} and Gitman $\&$
Kupriyanov \cite{gitman}. They consider the same problem as us,
but their strategy is different. In our opinion, their approach
fits much better the context of gauge field dynamics.

As Ludwig Faddeev noted during the Edward Witten's talk at the
\textbf{Mathematical Physics Conference: From XX To XXI
Century}\footnote{It took place at EPFL Laussane 16.-17.3.2009.},
\emph{quantization is not a science, quantization is an art}. Let
us believe that the quantization method proposed here will fit the
Ludwig's dictum and will be meaningful enough to be considered
artistic.

\section{Feynman Quantum Mechanics}\label{2}

According to Feynman \cite{feynman}, the probability amplitude of
the transition of the system from the space-time configuration
$(q_0,t_0)$ to another space-time configuration $(q_1,t_1)$ is
\begin{equation}\label{FeynmanI}
\mathbf{A}(q_1,t_1\,|\,q_0,t_0)=\frac{1}{\mathbf{N}}\int[\mathscr{D}\tilde{\gamma}]
\exp{\Bigl\{\frac{i}{\hslash}\int\limits_{\tilde{\gamma}}\hspace{-1mm}pdq-Hdt\Bigr\}}\,.
\end{equation}
Here the integral is taken over all paths (histories)
$\tilde{\gamma}(t)=(\tilde{q}(t),\tilde{p}(t),t)$ in the extended
phase space \footnote{For the record, the extended phase space is
locally described by the coordinates in the configuration space
$q^a$, the canonically conjugated momenta $p_a$ and the time
coordinate $t$ (the index $a$ runs from 1 to the number of degrees
of freedom $n$). One assumes the standard Poisson brackets
$\{p_a,q^b\}=-\delta_a^b$. In a more elevated language, the
extended phase space is the space $T^*M\times\mathbb{R}$, where
$M$ stands for the configuration space and $\mathbb{R}$ for
time.}, satisfying the conditions $\tilde{q}(t_0)=q_0$ and
$\tilde{q}(t_1)=q_1$.

The preexponential factor $1/\mathbf{N}$ in the expression
(\ref{FeynmanI}) serves just the normalization. To fix it properly
we impose two physical conditions on the transition amplitude.
First we introduce an integral condition that ensures that the
total probability is conserved,
\begin{subequations}\label{normalization}
\begin{equation}
\int
dq_1\,\overline{\boldsymbol{\mathbf{A}}}(q_1,t_1\,|\,q_0,t_0)\,\mathbf{A}(q_1,t_1\,|\,q'_0,t_0)
=\delta(q_0-q'_0)\,.
\end{equation}
This specifies the absolute value of $\mathbf{N}$. Then we add a
constraint that expresses the obvious fact that no evolution takes
place if the final time $ t_1$ approaches the initial time $t_0$,
\begin{equation}
\lim\limits_{t_1\rightarrow t_0}\mathbf{A}(q_1,t_1\,|\,q_0,t_0)=\delta(q_1-q_0)\,.
\end{equation}
\end{subequations}
This determines the phase of $\mathbf{N}$.

A miraculous consequence of the definition of the transition
amplitude (\ref{FeynmanI}) (not an additional requirement!) is
that it satisfies the evolutionary chain rule, or
Chapman-Kolmogorov equation,
\begin{equation}\label{Ch-K}
\mathbf{A}(q_1,t_1\,|\,q_0,t_0)=\hspace{-1mm}\int\hspace{-1mm}
dq'\mathbf{A}(q_1,t_1\,|\,q',t)\,\mathbf{A}(q',t\,|\,q_0,t_0)\,.\hspace{-1mm}
\end{equation}
The infinitesimal version of this formula is the celebrated
Schr\"{o}dinger equation.

Quantum states of the system are described by the square
integrable functions with the standard Hilbert space structure.
Physical observables are hermitian operators acting on such
functions. Given the state $\Psi_0(q)$ at the initial moment $t_0$
one is able to predict the state $\Psi_1(q)$ at any later moment
$t_1$ according to the formula
\begin{equation}\label{evolution}
\Psi_0(q)\ \rightsquigarrow\ \Psi_1(q)=\int
dq'\,\mathbf{A}(q,t_1\,|\,q',t_0)\Psi_0(q')\,.
\end{equation}
In what follows it is assumed that the above concept of states,
observables and quantum evolution is valid for the stringy
quantization of non-Lagrangian systems as well. The only new
element is a modified prescription for the evolutionary integral
kernel $\mathbf{A}(q_1,t_1\,|\,q_0,t_0)$. The same kinematical
prerequisites can be found also in other phenomenological
approaches \cite{CK}-\cite{geicke}.

\section{One step beyond Feynman}\label{3}

A possible step beyond the theory summarized above consists in the
elimination of the Hamiltonian function $H$ from formula
(\ref{FeynmanI}). The price to be paid is the replacement of the
path integration by the surface functional integration.

Our aim is to construct the amplitude for the transition between
$(q_0,t_0)$ and $(q_1,t_1)$ starting from the classical equations
of motion (and not from the Hamiltonian function which provides
them)
\begin{equation}\label{eq.}
\dot{q}^a=\frac{\partial H}{\partial p_a} \equiv \frac {p^a}m\,,
\quad \dot{p}_a=-\frac{\partial H}{\partial q^a} \equiv F_a\,.
\end{equation}
In the first set of equations, $m$ is the mass of the particle. We
restricted ourselves to the simplest case of one particle,
although it is trivial to generalize the theory to a system with
an arbitrary number of particles. Note also that if the particle
is unconstrained and we make use of Cartesian coordinates, the
momenta $p^a$ defined in (\ref{eq.}) reduce to $p_a$.

\begin{figure}
\begin{center}
\includegraphics[height=11.18cm,width=8.5cm]{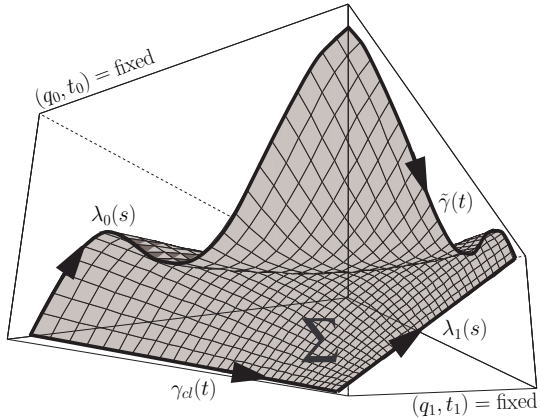}
\vspace{-4.5cm} \caption{Schematic picture of two auxiliary curves
$\lambda_0(s)$ and $\lambda_1(s)$ which connect the classical
history $\gamma_{cl}(t)$ with the given history
$\tilde{\gamma}(t)$ in the extended phase space. $\lambda$-curves
are located in the $n$-dimensional subspaces of the extended phase
space in which the momenta are varying while the coordinates and
time are kept fixed. The contour
$\partial\Sigma=\tilde{\gamma}-\lambda_1-\gamma_{cl}+\lambda_0$
forms a boundary for plenty of extended phase space surfaces. One
of them, denoted as $\Sigma$, is drawn in the figure.}
\label{umbilic}
\end{center}
\end{figure}

Suppose that there exists a unique classical trajectory in the
extended phase space $\gamma_{cl}(t)=(q_{cl}(t),p_{cl}(t),t)$,
connecting the points $(q_0,t_0)$ and $(q_1,t_1)$. Then we can
assign to any other trajectory
$\tilde{\gamma}(t)=(\tilde{q}(t),\tilde{p}(t),t)$, which enters
the path integral in (\ref{FeynmanI}), two auxiliary curves
\begin{eqnarray*}
\lambda_0(s)\hspace{-2mm}&=&\hspace{-2mm}\bigl(q_0,\pi(s),t_0\bigr),\
\mbox{where}\ \pi(0)=p_{cl}(t_0),\ \pi(1)=\tilde{p}(t_0);\\
\lambda_1(s)\hspace{-2mm}&=&\hspace{-2mm}\bigl(q_1,\phi(s),t_1\bigr),\
\mbox{where}\ \phi(0)=p_{cl}(t_1),\ \phi(1)=\tilde{p}(t_1).
\end{eqnarray*}
The curves are parameterized by the parameter $s\in[0,1]$ and live
in the momentum subsectors of the extended phase space with fixed
$(q_0,t_0)$ and $(q_1,t_1)$, see Figure \ref{umbilic}.
Using these definitions one can write \footnote{The integrals over
$\lambda_0$ and $\lambda_1$ give trivially zero contributions to
the contour integral, since on both $\lambda$'s $dq=0$ and
$dt=0$.}
\begin{equation}\label{loop}
\int\limits_{\tilde{\gamma}} p_adq^a-Hdt=\int\limits_{\gamma_{cl}}
p_adq^a-Hdt + \oint\limits_{\partial\Sigma} p_adq^a-Hdt\,,
\end{equation}
where
$\partial\Sigma=\tilde{\gamma}-\lambda_1-\gamma_{cl}+\lambda_0$ is
a contour in the extended phase space consisting of four curves
$\tilde{\gamma}(t)$, $\gamma_{cl}(t)$, $\lambda_0(s)$,
$\lambda_1(s)$. The first integral on the right hand side is the
classical action $\mathbf{S}_{cl}(q_1,t_1\,|\,q_0,t_0)$ (its
equivalent for the non-Lagrangian case will be specified later),
while the second integral can be rewritten as
$$
\oint\limits_{\partial\Sigma} p_adq^a-Hdt=\int\limits_{\Sigma}
dp_a\wedge\Bigl(dq^a-\frac{
\partial H}{\partial p_a}dt\Bigr)-\frac{\partial H}{\partial q^a}dq^a\wedge dt\,,$$
where $\Sigma$ is a surface spanning the contour $\partial\Sigma$,
i. e. a map from the parametric space
$(t,s)\in[t_0,t_1]\times[0,1]$ to the extended phase space,
$$
\Sigma: (t,s)\mapsto
\Sigma(t,s)\equiv\bigl(q^a(t,s),p_a(t,s),t(t,s)=t\bigr)\,,
$$
satisfying
\begin{equation*}
{\Sigma(t,0)=\gamma_{cl}(t) \atop \Sigma(t,1)=\tilde{\gamma}(t)\
\,}\ \ \ \ \mbox{and}\ \ \ \ {\Sigma(t_0,s)=\lambda_0(s)\atop
\Sigma(t_1,s)=\lambda_1(s)}\,.
\end{equation*}
The surface $\Sigma$ can be viewed as a worldsheet of a string,
therefore we will call the quantization method using such surfaces
``stringy.'' Partial derivatives of the Hamiltonian function
entering the integral over $\Sigma$ can be eliminated with the
help of the equations of motion (\ref{eq.}). By doing so, one
transforms (\ref{loop}) into the following form:
\begin{equation}\label{loop2}
\int\limits_{\tilde{\gamma}}
p_adq^a-Hdt=\mathbf{S}_{cl}(q_1,t_1\,|\,q_0,t_0)+\int\limits_{\Sigma}\Omega\,,
\end{equation}
where the two-form $\Omega$ is defined as
\begin{equation}\label{Omega}
\Omega=dp_a\wedge dq^a-\Bigl(\frac{p^a}{m}dp_a -
F_adq^a\Bigr)\wedge dt\,.
\end{equation}
This two-form is an object in the extended phase space and its
structure can be read out from the underlying equations of motion.
Note that for non-potential forces the expression $(p^a/m) dp_a -
F_adq^a$ does not reduce to $dH$, so that the two-form $\Omega$ is
not closed. This becomes essential in the next subsection.

\begin{figure}
\begin{center}
\includegraphics[height=11.18cm,width=8.5cm]{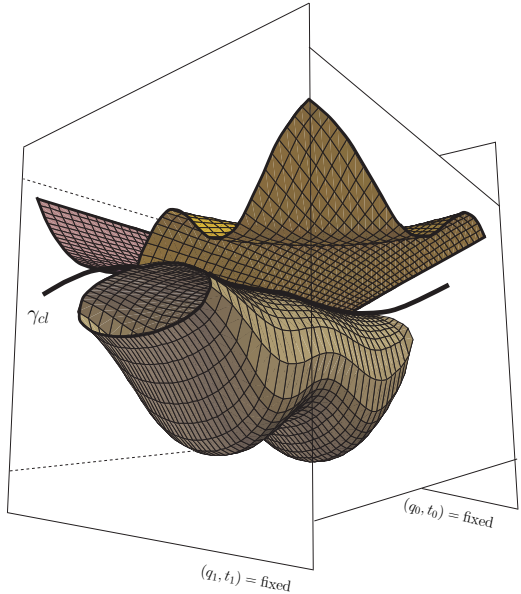}
\vspace{-2cm}
\caption{Four elements (two open, one closed and one shrunk to
$\gamma_{cl}$) of $\mathscr{U}_{\gamma_{cl}}$, anchored to the
given classical history $\gamma_{cl}$ in the extended phase space.
The two open surfaces belong to two different stringy classes
$\mathscr{U}_{(\tilde{\gamma},\gamma_{cl})}$, while the closed and
the shrunk surface are both elements of the same stringy class
$\mathscr{U}_{(\gamma_{cl},\gamma_{cl})}$. As the worldsheet
$\Sigma$ varies, the front and rear boundary curves, denoted in
the text $\lambda_0$ and $\lambda_1$, vary as well.}
\label{string}
\end{center}
\end{figure}

It is obvious that for a given pair of histories
$(\tilde{\gamma},\gamma_{cl})$ there exist infinitely many
$\Sigma$-surfaces such that
$\tilde{\gamma}-\gamma_{cl}\subset\partial\Sigma$. All of them
form a set which we will call
$\mathscr{U}_{(\tilde{\gamma},\gamma_{cl})}$ \footnote{The precise
definition is: suppose there are two histories
$\tilde{\gamma}_0(t)$ and $\tilde{\gamma}_1(t)$ in the extended
phase space whose projections $\mathit{Proj}(\tilde{\gamma}_0)$
and $\mathit{Proj}(\tilde{\gamma}_1)$ on the extended
configuration space connect $(q_0,t_0)$ with $(q_1,t_1)$. Then
$\mathscr{U}_{(\tilde{\gamma}_1,\tilde{\gamma}_0)}$ contains all
continuous and oriented surfaces $\Sigma$ in the extended phase
space such that
$$
\hspace{5mm}{\Sigma(t,s=0)=\tilde{\gamma}_0(t) \atop
\Sigma(t,s=1)=\tilde{\gamma}_1(t)}\ \ \ \mbox{and}\ \ \
{\mathit{Proj}\bigl(\Sigma(t=t_0,s)\bigr)=(q_0,t_0) \atop
\mathit{Proj}\bigl(\Sigma(t=t_1,s)\bigr)=(q_1,t_1)}\,.$$ If
$\tilde{\gamma}_0\neq\tilde{\gamma}_1$, the orientation of
$\Sigma$ is supposed to be such that
$\tilde{\gamma}_1-\tilde{\gamma}_0$ is a part of the boundary of
$\Sigma$. The $\mathit{Proj}$ection is simply the momentum
forgetting map, or to put it in a more elevated way, the canonical
projection from the cotangent bundle to the base manifold. A
special case is the set
$\mathscr{U}_{(\tilde{\gamma},\tilde{\gamma})}$, comprised of all
closed surfaces in the extended phase space containing
$\tilde{\gamma}$. In what follows, we consider any history
$\tilde{\gamma}$ as a degenerated (shrunk) closed surface, hence
$\tilde{\gamma}\in\mathscr{U}_{(\tilde{\gamma},\tilde{\gamma})}$
by definition.}. Since no $\Sigma$ is preferred and
$\int_\Sigma\Omega$ is only boundary dependent, it is natural to
average the exponent of (\ref{loop2}) over the whole stringy set
$\mathscr{U}_{(\tilde{\gamma},\gamma_{cl})}$. After doing so we
obtain the identity
$$
\exp{\Bigl\{\frac{i}{\hslash}\int\limits_{\tilde{\gamma}}p_adq^a-Hdt\Bigr\}}
=\frac{\mbox{\large{e}}^{\frac{i}{\hslash}\mathbf{S}_{cl}}}{\mathrm{N}_{\tilde{\gamma}}}\hspace{-1.4mm}
\int\limits_{\mathscr{U}_{(\tilde{\gamma},\gamma_{cl})}}\hspace{-3mm}[\mathscr{D}\Sigma]
\exp{\Bigl\{\frac{i}{\hslash}\int\limits_{\Sigma}\Omega\Bigr\}},
$$
where $\mathrm{N}_{\tilde{\gamma}}$ is the cardinality of the
stringy set $\mathscr{U}_{(\tilde{\gamma},\gamma_{cl})}$,
$\mathrm{N}_{\tilde{\gamma}}:=\#\mathscr{U}_{(\tilde{\gamma},\gamma_{cl})}$,
and $[\mathscr{D}\Sigma]$ is a functional integration measure
specified in Appendix \ref{B}. If no topology-related problems
arise in the extended phase space, the infinite constant
$\mathrm{N}_{\tilde{\gamma}}$ is independent of the history
$\tilde{\gamma}$. Taking all this into account we can rewrite
(\ref{FeynmanI}) as
\begin{equation}{\label{FeynmanIII}}\boxed{
\vspace{5mm} \hspace{1mm}\mathbf{A}(q_1,t_1\,|\,q_0,t_0)\,=\,
\frac{\mbox{\large{e}}^{\frac{i}{\hslash}\mathbf{S}_{cl}}}{\mathbf{N}}
\hspace{-1mm}\int\limits_{\mathscr{U}_{\gamma_{cl}}}^{^{^{}}}[\mathscr{D}\Sigma]
\exp{\Bigl\{\frac{i}{\hslash}\int\limits_{\Sigma}\Omega\Bigr\}},\hspace{2mm}
}\end{equation} where the set $\mathscr{U}_{\gamma_{cl}}$ over
which the functional integration is carried out contains all
strings in the extended phase space which are anchored to the
given classical trajectory $\gamma_{cl}$. Some surfaces from
$\mathscr{U}_{\gamma_{cl}}$ are depicted in Figure \ref{string}.
In (\ref{FeynmanIII}), the undetermined constant
$1/\mathrm{N}_{\tilde{\gamma}}$ was absorbed into the overall
preexponential factor $1/\mathbf{N}$ and the path integral over
$\tilde{\gamma}$'s was converted into the surface functional
integral, as promised earlier, using the identities
\begin{equation}\label{surface}
\int[\mathscr{D}\tilde{\gamma}]\hspace{-3.5mm}\int\limits_{\ \ \
\mathscr{U}_{(\tilde{\gamma},\gamma_{cl})}}
\hspace{-5mm}[\mathscr{D}\Sigma]\cdots\ =
\hspace{-5.5mm}\int\limits_{\ \ \
\bigcup_{\tilde{\gamma}}\mathscr{U}_{(\tilde{\gamma},\gamma_{cl})}}
\hspace{-6.7mm}[\mathscr{D}\Sigma]\cdots\ =
\hspace{-1.6mm}\int\limits_{\mathscr{U}_{\gamma_{cl}}}\hspace{-1.0mm}[\mathscr{D}\Sigma]\cdots\
\,.
\end{equation}

%\newpage

%\enddocument

\section{One more step beyond -- non-Lagrangian systems}\label{4}

In case the Hamiltonian function $H$ is given, the formula
(\ref{FeynmanIII}) for the propagator is equivalent to the Feynman
formula (\ref{FeynmanI}) we have started with. The surface
functional integral (\ref{FeynmanIII}) is however completely free
of $H$ and requires just the knowledge of the classical equations
of motion. This observation allows us to postulate
(\ref{FeynmanIII}) as a quantization tool in situations in which
one cannot use the standard Hamiltonian approach. We just have to
relax the requirement of the closedness of the two-form $\Omega$,
following from the definition of $\Omega$ for Hamiltonian systems.
This relaxation is what is hidden behind the slightly provocative
phrase \emph{one more step beyond} in the title of this
subsection.

The problematic part here is the definition of the classical
action $\mathbf{S}_{cl}(q_1,t_1\,|\,q_0,t_0)$ for non-Lagrangian
systems. Later we will see that in some specific situations this
quantity can be read out from the structure of the surface
integral
$$
\int\limits_{\mathscr{U}_{\gamma_{cl}}}[\mathscr{D}\Sigma]\exp{\Bigl\{\frac{i}
{\hslash}\int\limits_{\Sigma}\Omega\Bigr\}}\,.
$$
Another possibility which comes to mind is to use the integrating
factor of the generally non-closed two-form $\Omega$. This means
that one will look for a function $f(q,p,t)$ on the extended phase
space such that $d(f\Omega)=0$. If the structure of the dynamical
equations allows for such a function (i.e. the forces satisfy the
Helmholtz condition) then we can define a local auxiliary
Hamiltonian $H_{aux}$ such that $f\Omega=d(p_adq^a-H_{aux}dt)$.
Having $H_{aux}$ we can define the auxiliary classical action
$\mathbf{S}_{aux}$. The alternative approach which tries to
identify $\mathbf{S}_{cl}$ with $\mathbf{S}_{aux}$ has, however,
several substantial disadvantages. First of all, the procedure of
finding $f$ (and subsequently $H_{aux}$ and $\mathbf{S}_{aux}$) is
highly ambiguous. Furthermore, $\mathbf{S}_{aux}$ lacks symmetries
that were originally present in the equations of motion. Because
of these findings we do not follow this strategy hereinafter.

From the physical point of view the dynamical equations seem to be
more fundamental than their compact but ambiguous precursors,
Hamiltonian and/or Lagrangian function, see \cite{guys}. Since
(\ref{FeynmanIII}) requires just the knowledge of the equations of
motion, it determines the transition amplitude in a completely new
way.

Our proposal for the transition amplitude has appeared here out of
thin air. Actually, we were rewriting (\ref{FeynmanI}) in terms of
stringy surfaces and then, when realizing that the integrated
function can be written without any reference to the Hamiltonian,
we postulated formula (\ref{FeynmanIII}) to be valid in general.
However, this is not the whole story. The stringy functional
quantization can be motivated also by the stringy variational
principle. Having the dynamical equations and initial and final
endpoints, one can form $\mathscr{U}_{\gamma_{cl}}$ and $\Omega$.
Then using these objects one can introduce the stringy action
functional
\begin{equation}\label{action}
\mathscr{S}:\mathscr{U}_{\gamma_{cl}}\rightarrow \mathbb{R}\,,\ \ \ \ \
\Sigma\mapsto \mathscr{S}(\Sigma):=\int\limits_{\Sigma}\Omega\,.
\end{equation}
This is a variational problem with varying boundaries, therefore
the total variation has two terms. First one specifies the
boundary and determines the initial equations of motion for
$\gamma_{cl}$. The second one specifies the bulk of the stationary
world-sheet $\Sigma$ (which turns out to be shrunk into
$\gamma_{cl}$ itself). Moreover, one immediately verifies that in
the special case when $\Omega=d(p_adq^a-Hdt)$, the stringy
variational principle reduces (up to an additive constant) to the
celebrated Hamilton least action principle. Of course, many
subtleties were omitted here, but all of them can be found in
\cite{kochan}, or in Appendix \ref{A}. Thus, our variational
principle enables us to perform the limit $\hslash\rightarrow 0$,
in which we recover the original classical dynamics as required.

\section{Quantization of friction force systems}\label{4.1}

To examine the functionality of the proposed quantization method
let us first analyze the simplest friction force system. It
consists of a particle with unit mass, moving in one dimension
under the action of the conservative force $F=-dU/dq$ and the
friction force $-\kappa\dot{q}^{A}$ \footnote{Strictly speaking,
the expression for the friction force can be used only for the
motion with increasing $q$. The universally applicable expression
is $-\kappa$\hskip 0.5mm sgn$(\dot{q})|\dot{q}|^{A}$.}. Thus,
$$
\ddot{q}=-\kappa\dot{q}^{A}+F\, \Rightarrow\,
\Omega=d\bigl(pdq-\tfrac{1}{2}p^2dt-Udt\bigr)-\kappa
p^{A}\,dq\wedge dt\,.
$$
In this example the surface functional integral can be calculated
explicitly (for more detail see \cite{kochan}, or Appendix
\ref{B}). In the course of calculation, the surface functional
integral in the extended phase space reduces to path integral in
the configuration space,
\begin{eqnarray}\label{disi}
&&\hspace{-4mm}\int\limits_{\mathscr{U}_{\gamma_{cl}}}[\mathscr{D}\Sigma]\exp{\Bigl\{\frac{i}{\hslash}
\int\limits_{\Sigma}\Omega\Bigr\}}\propto\nonumber\label{dissipative}\\
&&\propto\exp{\Bigl\{-\frac{i}{\hslash}\int\limits_{t_0}^{t_1}\bigl(\tfrac{1}{2}\dot{q}_{cl}^2-U(q_{cl})-
\kappa q_{cl}p^{A}_{cl}\bigr)dt\Bigr\}}\times\nonumber\\
&&\times\int[\mathscr{D}q]\exp{\Bigl\{\frac{i}{\hslash}\int\limits_{t_0}^{t_1}\bigl(\tfrac{1}{2}\dot{q}^2-U(q)-
\kappa qp^{A}_{cl}\bigr)dt\Bigr\}}\,.
\end{eqnarray}
This combined with the formula (\ref{FeynmanIII}) suggests that it
will be convenient to define the classical action as \footnote{The
prefix $A$ in the subscript $Acl$, as well as in the subscript
$Astr$ introduced later, refers to the power in the expression for
the friction force.}
\begin{equation}\label{A-action}
\mathbf{S}_{Acl}(q_1,t_1|q_0,t_0)=\int\limits_{t_0}^{t_1}\bigl(\tfrac{1}{2}
\dot{q}_{cl}^2-U(q_{cl})-\kappa q_{cl}p^{A}_{cl}\bigr)dt\,.
\end{equation}
When doing so, we obtain a controlled cancelation of
$\exp\{\frac{i}{\hslash}\mathbf{S}_{Acl}\}$ with the
preexponential factor arising from (\ref{disi}). The final
probability amplitude then assumes a compact and reasonable form:
\begin{eqnarray}\label{propagator}
&&\hspace{-1cm}\mathbf{A}_{Astr}(q_1,t_1|q_0,t_0)=\nonumber\\
&&\hspace{-1cm}=\frac{1}{\mathbf{N}_{Astr}}\int[\mathscr{D}q]\exp{\Bigl\{\frac{i}{\hslash}
\hspace{-0.5mm}\int\limits_{t_0}^{t_1}\hspace{-0.5mm}\bigl(\tfrac{1}{2}\dot{q}^2-\hspace{-0.2mm}
U(q)-\hspace{-0.2mm}\kappa qp^{A}_{cl}\bigr)dt\Bigr\}}\,.
\end{eqnarray}
The preexponential factor $1/\mathbf{N}_{Astr}$ can in principle
be obtained by subjecting $\mathbf{A}_{Astr}(q_1,t_1|q_0,t_0)$ to
the conditions (\ref{normalization}).

Let us explain why one should consider the propagator formula
(\ref{propagator}) reasonable. In the path integral above there
appears an unconventional ``external source'' term $-\kappa
qp^{A}_{cl}$. Its appearance guarantees that the quantum dynamics
governed by $\mathbf{A}_{Astr}(q_1,t_1|q_0,t_0)$ transforms into
classical mechanics in the limit $\hslash \to 0$. This follows
from the simple fact that the unique solution of the saddle point
equation for the path integral (\ref{propagator}),
$$
\ddot{q}=-\frac{dU}{dq}-\kappa\dot{q}^{A}_{cl}\,,
$$
which satisfies the given initial and final conditions
$q(t_0)=q_0$ and $q(t_1)=q_1$, is the classical trajectory
$q_{cl}(t)$.

The external source term in (\ref{propagator}) breaks the validity
of the Chapman-Kolmogorov (memoryless) equation \footnote{The
external source in the functional integral (\ref{propagator}) can
be viewed as a sort of additional time-dependent potential,
therefore one would expect that the Chapman-Kolmogorov equation
(\ref{Ch-K}) will be satisfied. However, this argument is not
applicable in the present situation. The point is that if we merge
two paths \emph{KL} and \emph{LM} into a new path \emph{KM}, then
the time-dependent potentials $-\kappa qp^A_{cl-KL}(t)$ and
$-\kappa qp^A_{cl-LM}(t)$ differ, for a general position of
$\emph{L}$, from the potential $-\kappa qp^A_{cl-KM}(t)$ in the
corresponding time (sub)intervals,
$$
-\kappa qp^A_{cl-KM}(t)\neq \Bigl\{{-\kappa qp^A_{cl-KL}(t)\
\mbox{for}\ t\in[t_K,t_L] \atop -\kappa qp^A_{cl-LM}(t)\ \mbox{for}\ t\in[t_L,t_M]}
$$
Since this differs substantially from the standard case, the
argument about the validity of the Chapman-Kolmogorov equation
cannot be used.}. From the physical point of view it is a desired
phenomenon. The microscopic origin of friction is some
environmental interaction. This, however, was not accounted for
here explicitly. What has been considered is some effective
(macroscopic, phenomenological) interaction emerging on the
classical level only. Microscopically the system is a part of a
larger system and hence it should be affected by the memory
effect.

It is not difficult to compute
$\mathbf{A}_{Astr}(q_1,t_1|q_0,t_0)$ for the general power $A$ and
the potentials $U(q)=0$ and $U(q)=\tfrac{1}{2}\omega^2 q^2$. In
both cases we can carry out the path integration in
(\ref{propagator}) explicitly to obtain
\begin{eqnarray}\label{propagator1}
&&\hspace{-1cm}\mathbf{A}_{Astr}(q_1,t_1|q_0,t_0)=
\nonumber\\
&&=\frac{1}{\mathbf{N}_{Astr}}\exp\left\{\frac{i}{\hslash}
\mathbf{S}_{Acl}(q_1,t_1|q_0,t_0)\right\}\,,
\end{eqnarray}
where the non-Lagrangian action
$\mathbf{S}_{Acl}(q_1,t_1|q_0,t_0)$ is given by the expression
(\ref{A-action}). An open problem is to determine the
preexponential factor $1/\mathbf{N}_{Astr}(t_1-t_0,q_1,q_0)$ for
all powers $A$ except for $A = 1$ (at least the author is
incapable to do that). If $A\neq 1$, the preexponential factor is
apparently dependent not only on the time difference $t_1-t_0$,
but also on the endpoint positions $q_0$ and $q_1$. This
hypothesis is supported by the results obtained in
\cite{stuckens-kobe}, where the path integral for a
non-conservative force quadratic in velocity is calculated. The
normalization prefactor which appears there is explicitly
dependent on the endpoints positions as well as on $t_1-t_0$.

In what follows we will restrict ourself to the case when the
power $A$ is equal to 1. This simplified setup enables us to
compare the stringy approach with the Caldirola-Kanai, Bateman and
Kostin approaches \footnote{An exhaustive list of various
quantization techniques applicable in the presence of linear
dissipation, including those mentioned in the introduction, can be
found in \cite{dekker} and \cite{razavy}.}.

\subsection{Stringy \emph{versus} Caldirola-Kanai approach}

\begin{figure}
\includegraphics[height=11.18cm,width=8.5cm]{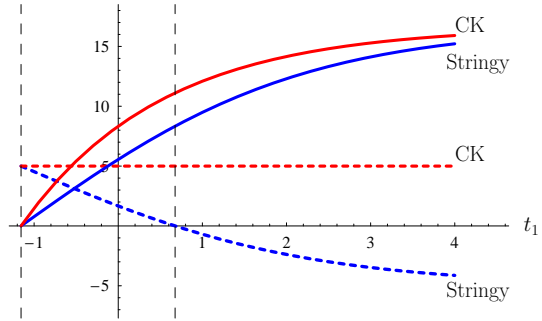}
\vspace{-5.5cm}
\caption{Expectation values of position (solid color lines) and
momentum (dashed color lines) as functions of time for the
Gaussian wave packet (the best fit for the moving classical
particle), drawn for the stringy and CK propagators (\ref{damp.A})
and (\ref{damp.B}), respectively. Initial wave packet
characteristics are $\langle q\rangle(t_0)=0\,\mathrm{m}$ and
$\langle p\rangle(t_0)=5\,\mathrm{m\,s^{-1}}$, and the friction
constant $\kappa$ is set to $0.6\,\mathrm{s^{-1}}$. One realizes
that at the time $t_1-t_0=\ln{3}\ \kappa^{-1}$ the stringy mean
momentum becomes zero. Classically, for $t_1-t_0>\kappa^{-1}$ the
physical relevancy of the solution breaks down. This explains the
time bound on the applicability of stringy propagator
(\ref{damp.A}).} \label{CK-characteristics}
\end{figure}

The dissipative system under consideration is weakly
non-Lagrangian. This means that $d\Omega\neq 0$, but $\Omega$
possesses a local integrator $f\neq 0$. Let us define the
auxiliary Lagrangian function (one of plenty) as
\begin{align}
L_{CK}(q,t)&=e^{\kappa
t}\Bigl[\tfrac{1}{2}\dot{q}^2-U(q)\Bigr].\label{L-CK}
%A&=2   & L_{HNT}(q,\dot{q})&=e^{2\kappa q}\,\frac{\dot{q}^2}{2}-\int^q\hspace{-2mm}d\tilde{q}\,
%e^{2\kappa\tilde{q}}\,\frac{dU(\tilde{q})}{d\tilde{q}},\label{L-HNT}
\end{align}
This Lagrangian is usually called CK-Lagrangian after its
inventors P. Caldirola and E. Kanai \cite{CK}. Henceforward a
damped free particle is considered only, i.e. the potential energy
$U(q)$ is supposed to be zero. It is immediately clear that the
transition amplitude computed from (\ref{propagator1}) differs
from the CK-amplitude
\begin{eqnarray*}
\mathbf{A}_{CK}(q_1,t_1|q_0,t_0)\hspace{-1mm}&=&\hspace{-1mm}\dfrac{1}{\mathbf{N}_{CK}}
\int[\mathscr{D}q]\exp{\Bigl\{\frac{i}{\hslash}\int\limits_{t_0}^{t_1}\hspace{-1mm}L_{CK}(q,t)dt\Bigr\}}\\
&=&\hspace{-1mm}\dfrac{1}{\mathbf{N}_{CK}}\exp\{\tfrac{i}{\hslash}\,\mathbf{S}_{CK}(q_1,t_1|q_0,t_0)\}
\,.
\end{eqnarray*}
Indeed, the stringy and CK actions which enter the corresponding
transition amplitudes are different,
\begin{eqnarray*}
&&\mathbf{S}_{1cl}=\frac{\kappa}{4}(q_1-q_0)\frac{(q_0+3q_1)\mathrm{e}^{-\kappa
t_1}-(q_1+3q_0)\mathrm{e}^{-\kappa t_0}}
{\mathrm{e}^{-\kappa t_0}-\mathrm{e}^{-\kappa t_1}},\\
&&\mathbf{S}_{CK}=\frac{\kappa}{2}\,\frac{(q_1-q_0)^2}{\mathrm{e}^{-\kappa
t_0}-\mathrm{e}^{-\kappa t_1}}.
\end{eqnarray*}
Since both actions are quadratic functions of the endpoints $q_0$
and $q_1$, and since we require that the total probability is
conserved, the preexponential factors for both transition
amplitudes can be computed from the Van Vleck formula
$$%\begin{equation}\label{van-vleck}
\frac{1}{\mathbf{N}}=\sqrt{\frac{i}{2\pi\hslash}\,\frac{\partial^2 \mathbf{S}}{\partial q_1\,\partial q_0}}\,.
$$%\end{equation}
Using this formula we arrive at the normalized transition
amplitudes
\begin{eqnarray}
\hspace{-7mm}\mathbf{A}_{1str}(q_1,t_1|q_0,t_0)\hspace{-1.8mm}&=&\hspace{-1.8mm}
\sqrt{\frac{\kappa}{4\pi i \hslash\tanh\tfrac{\kappa}{2}(t_1-t_0)}}\,
\mbox{\large{e}}^{\frac{i}{\hslash}\mathbf{S}_{1cl}},\label{damp.A}\\
\hspace{-7mm}\mathbf{A}_{CK}(q_1,t_1|q_0,t_0)\hspace{-1.8mm}&=&\hspace{-1.8mm}
\sqrt{\frac{\kappa}{2\pi i \hslash\,(\mathrm{e}^{-\kappa t_0}-\mathrm{e}^{-\kappa t_1})}}\,
\mbox{\large{e}}^{\frac{i}{\hslash}\mathbf{S}_{CK}},\label{damp.B}
\end{eqnarray}
which trivially satisfy (\ref{normalization}). Moreover, one
immediately verifies that in the frictionless limit
$\kappa\rightarrow 0$ both transition amplitudes coincide with the
free particle propagator. A short inspection shows that the
stringy propagator depends on $t_0$ and $t_1$ only through
$t_1-t_0$, but this is not the case for the CK propagator. The
same observation holds for $\mathbf{S}_{1cl}$ and
$\mathbf{S}_{CK}$. This is a typical feature for all auxiliary
actions defined in terms of integrators of $\Omega$, see \cite{all
others II}. We pay attention to this point because the classical
equation of motion $\ddot{q}=-\kappa\dot{q}$ which we have started
with is invariant with respect to time translations, and one would
naturally expect the same invariance on the quantum level. Since
only $\mathbf{A}_{1str}(q_1,t_1|q_0,t_0)$ has this feature, we
obtain an efficient argument for the stringy functional integral
(\ref{FeynmanIII}) when compared to the CK one. The evolution of
Gaussian wave packets
$$
\Psi_0(q)\propto\mbox{\large{e}}^{-q^2+\tfrac{i}{\hslash}p_0q}\,\rightsquigarrow\,
\Psi_1(q)\propto\hspace{-1mm}\int\hspace{-1mm} dq'\mathbf{A}(q,t_1\,|\,q',t_0)\Psi_0(q')\,,
$$
whose dynamics is governed by (\ref{damp.A}) and (\ref{damp.B}),
is visualized in Figure \ref{CK-characteristics}. For more detail,
see \cite{kochan2}.

From the figure we can see that the stringy averaged momentum
$\langle p\rangle$ becomes negative (i.e. meaningless) for
$t_1-t_0>\ln{3}\ \kappa^{-1}$. Let us try to explain this
peculiarity. It is well known that the relevancy of the classical
solution $q_{cl}(t)$ of $\ddot{q}=-\kappa\dot{q}$ breaks down when
the time $t_1-t_0$ exceeds the relaxation time $\kappa^{-1}$.
Since $q_{cl}$ is used in the derivation of the formula
(\ref{damp.A}) for the stringy propagator, its applicability is
automatically restricted as well. %Moreover, quantum
%fluctuations can pull $\langle p\rangle$ down to zero, however,
%classically one would expect $\langle p\rangle\simeq p_0e^{-1}$.
After this restriction is taken into account, the stringy
evolution of $\langle p\rangle$ becomes more acceptable from the
physical point of view then the evolution in the CK approach.

\subsection{Stringy \emph{versus} Bateman approach}

The key element of the Bateman(-Morse-Feshbach) approach
\cite{bateman} are new subsidiary degrees of freedom introduced in
addition to the initial ones, which are amplified rather than
damped; thus, they evolve according to the time reversed dynamical
equations. In our case we have:
$$
\ddot{q}=-\kappa\dot{q}\ \ \mbox{(damped) \ AND}\ \
\ddot{Q}=+\kappa\dot{Q} \ \ \mbox{(amplified)}\,.
$$
These equations of motion can be derived from the least action
principle with the quadratic and time independent
Bateman(-Morse-Feshbach) Lagrangian:
\begin{equation}\label{Bateman lagrangian}
L_B(q,\dot{q},Q,\dot{Q})=\dot{q}\dot{Q}+\frac{\kappa}{2}(q\dot{Q}-Q\dot{q})\,.
\end{equation}
The canonical as well as path integral quantization of the theory
encounters various difficulties because non-normalizable states of
the amplified system must be employed \cite{ghosh-hasse}. This
problem will be discussed later, when the treatment of the
auxiliary $Q$-degrees of freedom in this approach will be
described. The path integral evaluation of the transition
probability amplitude based on $L_B$ is straightforward
\cite{chetouani}. The result is:
\begin{eqnarray}
&&\hspace{-1cm}\mathbf{A}_{B}(q_1,Q_1,t_1|q_0,Q_0,t_0)=\frac{\kappa}
{4\pi\hslash\sinh{\frac{\kappa T}{2}}}\times \nonumber\label{bateman propagator}\\
&&\hspace{-5mm}\times\exp\Bigl\{\frac{i}{\hslash}\alpha\bigl[Q_1(q_1-q_0\beta_{-})+
Q_0(q_0-q_1\beta_{+})\bigr]\Bigr\}\,,
\end{eqnarray}
where
$$
\alpha=\frac{\kappa}{2}\,\frac{1}{\tanh{\frac{\kappa T}{2}}}\,,\ \
\ \beta_{\pm}=\frac{\mathrm{e}^{\pm\frac{\kappa
T}{2}}}{\cosh{\frac{\kappa T}{2}}}\,, \ \ \ T=t_1-t_0\,.
$$
Our aim, however, is to find the transition amplitude for the
damped (sub)system only. In order to obtain it we must project out
the nonphysical $Q$-degrees of freedom. When using the standard
formula
$$
\mathbf{A}_{B}^{\mathrm{trial}}(q_1,t_1|q_0,t_0)=\int dQ A_{B}(q_1,Q,t_1|q_0,Q,t_0)\,,
$$
we do not reproduce the free particle propagator for $\kappa=0$ as
desired. To overcome this trouble the following ``repairing
prescription'' is introduced (for more detail see
\cite{nemes-toledo piza}): the amplitude for the damped
(sub)system to pass from $|q_0\rangle$ to $|q_1\rangle$ within the
time $T$ is equal to the amplitude for the whole system to pass
from the nonphysical state $|q_0\rangle|\Psi_{-}\rangle$ to the
nonphysical state $|q_1\rangle|\Psi_{+}\rangle$,
\begin{eqnarray}
&&\hspace{-1cm}\mathbf{A}_{B}^{\mathrm{eff}}(q_1,t_1|q_0,t_0):=
\mbox{Amp}\Bigl(|q_0\rangle|\Psi_{-}\rangle\rightarrow|q_1\rangle|\Psi_{+}\rangle\Bigr)=\nonumber\label{remedy}\\
&&\hspace{-1cm}=\int dQdQ'\,
\overline{\Psi}_{+}(Q')\,A_{B}(q_1,Q',t_1|q_0,Q,t_0)\,\Psi_{-}(Q)\,,
\end{eqnarray}
where the non-normalizible $Q$-system states $|\Psi_{\pm}\rangle$
are chosen as
$$
\Psi_{\pm}(Q)=\sqrt{i\cosh{\tfrac{\kappa}{2}T}\sqrt{\frac{\mp 2i\alpha}{\pi\hslash}}}
\exp\bigl\{\pm \frac{i}{\hslash}\alpha Q^2\bigr\}\,.
$$
Substituting (\ref{bateman propagator}) into (\ref{remedy}) we
obtain the following effective propagator for the damped
(sub)system:
\begin{eqnarray}
&&\hspace{-1cm}\mathbf{A}_{B}^{\mathrm{eff}}(q_1,t_1|q_0,t_0):=\sqrt{\frac{\kappa}
{4\pi i\hslash\tanh{\frac{\kappa}{2}T}}}\times \nonumber\label{damp.C}\\
&&\hspace{-5mm}\times\exp\Bigl\{\frac{i}{\hslash}\frac{\alpha}{4}\Bigl[\bigl(q_1-q_0
\beta_{-}\bigr)^2+\bigl(q_0-q_1\beta_{+}\bigr)^2\Bigr]\Bigr\}\,.
\end{eqnarray}
The form of the auxiliary states $|\Psi_{+}\rangle$ and $|\Psi_{-}\rangle$ guarantees that the effective Bateman
propagator for the damped (sub)system satisfies the normalization condition (\ref{normalization}).

\begin{figure}
\includegraphics[height=11.18cm,width=8.5cm]{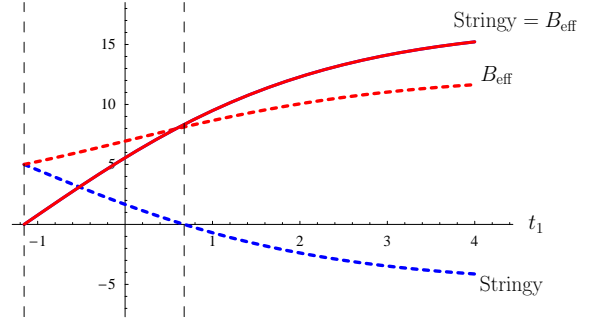}
\vspace{-5.5cm}
\caption{Expectation values of the position and momentum as
functions of time for the Gaussian wave packet, drawn for the
stringy and effective Bateman propagators (\ref{damp.A}) and
(\ref{damp.C}), respectively. The initial wave packet
characteristics and the meaning of solid and dashed color lines
are the same as in Figure \ref{CK-characteristics}.}
\label{Bateman-characteristics}
\end{figure}

The Bateman effective propagator (\ref{damp.C}) depends only on
the time difference $T$, which favors it in comparison with the
CK-propagator (\ref{damp.B}). On the other hand, the artificial
``repairing procedure'' involving nonphysical states discredits
the Bateman approach with respect to the stringy one. The Gaussian
wave packet characteristics obtained from (\ref{damp.A}) and
(\ref{damp.C}) are drawn in Figure \ref{Bateman-characteristics}.
As seen from the figure, the stringy quantization possesses again
better qualitative features than the approach to which we have
compared it.

\subsection{Stringy \emph{versus} Kostin approach}

\begin{figure}
\includegraphics[height=11.18cm,width=8.5cm]{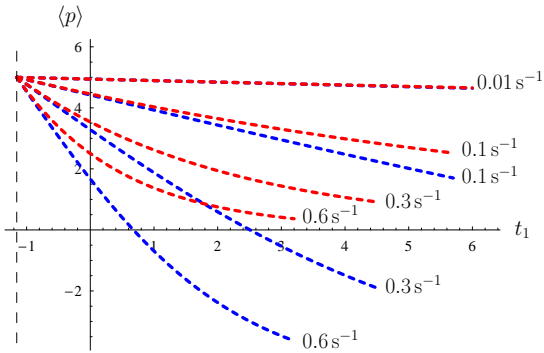}
\vspace{-5.5cm}
\caption{Expectation value of the momentum as a function of time
for an eigenstate of initial momentum
$\Psi_{p_0}(q,t_0)\propto\exp\{\frac{i}{\hslash}qp_0\}$, drawn for
the stringy propagator (\ref{damp.A}) (blue dashed lines) and the
propagator given by the Kostin-Schr\"{o}dinger equation
(\ref{Kostin}) (red dashed lines), respectively. The values of the
friction constant $\kappa$ are indicated next to the curves. The
initial momentum is $p_0=5\,\mathrm{m\,s^{-1}}$.}
\label{Kostin-characteristics}
\end{figure}

A possible incorporation of the classical dynamics into the
quantum one can be obtained by using Hei\-sen\-berg equations for
the position and momentum operators \footnote{The ordering problem
and its consequences are not discussed here.}:
$$
\dot{\hat{q}}=\hat{p}\,,\ \ \ \dot{\hat{p}}=F(\hat{q},\hat{p})\,,
$$
where $\hat{q}$ and $\hat{p}$ obey
$\bigl[\hat{q},\hat{p}\bigr]=i\hslash$ for all $t$. In the special
case when $F(\hat{q},\hat{p})=-dU(\hat{q})/dq-\kappa\hat{p}$,
Kostin found an equivalent description of the system by the
Schr\"{o}dinger equation \cite{kostin}:
\begin{equation}\label{Kostin}
i\hslash\partial_t\Psi(q,t)=\Bigl[-\frac{\hslash^2}{2}\frac{d^2}{dq^2}
+U(q)+\hslash K[\Psi]\Bigr]\Psi(q,t)\,,
\end{equation}
where $K[\Psi]$ is the $\Psi$-dependent Kostin potential defined
as
$$
K[\Psi]=i\frac{\kappa}{2}\Bigl[\ln{\frac{\overline{\Psi}(q,t)}{\Psi(q,t)}}-
\int\hspace{-1.0mm} dq'\,\overline{\Psi}(q',t)\ln{\frac{\overline{\Psi}(q',t)}{\Psi(q',t)}}\Psi(q',t)\Bigr].
$$
This is known as Kostin-Schr\"{o}dinger(-Langevin) equation. In
the Kostin's paper a complete set of solutions of this equation in
the special case of free particle ($U(q)=0$) is presented. The set
consists of the states $\Psi_{p_0}(q,t)$ labeled by the continuous
quantum number $p_0$, the initial momentum of the particle. The
state $\Psi_{p_0}(q,t)$ starts as a momentum eigenstate with the
momentum $p_0$, and remains the momentum eigenstate also later,
but with the decreasing momentum,
$$
\hat{p}\Psi_{p_0}(q,t)=p_0\mathrm{e}^{-\kappa(t-t_0)}\Psi_{p_0}(q,t)\,.
$$
From the ensemble of non-localized Kostin's states one would like
to form normalized wave packet solutions, localized in
configuration as well as momentum space. This is, however,
impossible since the Kostin-Schr\"{o}dinger equation is nonlinear;
thus, the Kostin approach belongs to nonlinear quantum mechanics
and lacks superposition principle. In Figure
\ref{Kostin-characteristics} we plot the expectation value of the
momentum for the Kostin state $\Psi_{p_0}(q,t_1)$ and the stringy
evolved state $\Psi(q,t_1)=\int dq'
\mathbf{A}_{1str}(q,t_1|q',t_0)\Psi_{p_0}(q',t_0)$ for various
values of $\kappa$ \footnote{Since the traveling wave
$|p_0\rangle$ is not a square integrable function, we evolved by
$\mathbf{A}_{1str}(q,t_1|q',t_0)$ a modified (normalized) state
$|\Psi_\epsilon\rangle=\int
dp\,\mathrm{e}^{-\epsilon(p-p_0)^2}|p\rangle$ ($\epsilon>0$).
After the stringy evolution of $|\Psi_\epsilon\rangle$, the
expectation value of the momentum was computed and the
regularization parameter $\epsilon$ was set to zero.}. As $\kappa$
tends to zero, the stringy and Kostin results match each other.

\section{Quantization of the Douglas system}\label{4.2}

After we have dealt with the weak non-Lag\-ran\-geanity, let us
consider the simplest strongly non-Lagrangian system. It was
proposed by Jesse Douglas (one of the two winners of the first
Fields Medals awarded in 1936) when studying the inverse problem
of variational calculus \cite{douglas}. The system is governed by
the following dimensionless dynamical equations:
\begin{equation}\label{eqs.}
\ddot{x}=-\dot{y},\ \ \ddot{y}=-y\ \ \ \Longleftrightarrow\ \ \ {p_x=\dot{x},
\atop p_y=\dot{y},}\ \ \ {\dot{p}_x=-p_y, \atop \dot{p}_y=-y\,.}
\end{equation}
A quick calculation shows that the associated two-form
$$
\Omega=d\bigl(p_xdx+p_ydy-\frac{p^2_x}{2}dt-\frac{p^2_y}{2}dt-\frac{y^2}{2}dt\bigr)-p_ydx\wedge dt
$$
does not possess a non-trivial local integrator $f$. Consequently,
equations (\ref{eqs.}) cannot be obtained as Euler-Lagrange
(Hamilton) equations.

From the standard point of view, this curious situation is
stalemate: \emph{no Lagrangian $\Rightarrow$ no Quantum
Mechanics}. However, the surface integral method offers a way to
overcome this deadlock. Using the stringy functional integration
in the extended phase space we get
\begin{eqnarray}
&&\hspace{-4mm}\int\limits_{\mathscr{U}_{\gamma_{cl}}}\hspace{-1mm}[\mathscr{D}\Sigma]
\exp{\Bigl\{\frac{i}{\hslash}\int\limits_{\Sigma}\Omega\Bigr\}}\propto\nonumber\label{douglas}\\
&&\hspace{-1cm}\propto
\exp{\Bigl\{-\frac{i}{\hslash}\int\limits_{t_0}^{t_1}
\bigl(\tfrac{1}{2}\dot{x}_{cl}^2+\tfrac{1}{2}\dot{y}_{cl}^2-\tfrac{1}{2}y_{cl}^2-x_{cl}p_{ycl}
\bigr)dt\Bigr\}}\hspace{-1mm}\times\nonumber\\
&&\hspace{-1cm}\times\hspace{-1mm}\int[\mathscr{D}x\,\mathscr{D}y]\exp{\Bigl\{\frac{i}{\hslash}
\int\limits_{t_0}^{t_1}\hspace{-1mm}\bigl(\tfrac{1}{2}\dot{x}^2+\tfrac{1}{2}\dot{y}^2-\tfrac{1}{2}y^2-xp_{ycl}
\bigr)dt\Bigr\}}.
\end{eqnarray}
Here, obviously, $(x_{cl}(t),p_{xcl}(t))$ and
$(y_{cl}(t),p_{ycl}(t))$ stand for the classical solutions of
(\ref{eqs.}) matching the initial and final endpoints for which
the quantum transition probability amplitude is sought for,
$x_{cl}(t_0)=x_0$, $y_{cl}(t_0)=y_0$ and $x_{cl}(t_1)=x_1$,
$y_{cl}(t_1)=y_1$. In the classical limit $\hslash\rightarrow 0$
we want the stringy propagator (\ref{FeynmanIII}) to maintain the
properties of the original dynamical system. This guides us to
define:
$$
\mathbf{S}_{Dcl}(x_1,y_1,t_1|x_0,y_0,t_0)\hspace{-0.5mm}=\hspace{-1.5mm}\int\limits_{t_0}^{t_1}\hspace{-1mm}
\bigl(\tfrac{1}{2}\dot{x}_{cl}^2+\tfrac{1}{2}\dot{y}_{cl}^2-\tfrac{1}{2}y_{cl}^2-x_{cl}p_{ycl}\bigr)dt\,.
$$
The exponential $\exp\{\frac{i}{\hslash}\mathbf{S}_{Dcl}\}$ of the
classical action defined in such a way cancels the prefactor in
(\ref{douglas}). Finally we arrive at the following transition
amplitude for the Douglas system:
\begin{eqnarray}
&&\hspace{-4mm} \mathbf{A}_{D}(x_1,y_1,t_1|x_0,y_0,t_0)=\frac{1}{\mathbf{N}_D}\hspace{-1mm}\times\nonumber\\
&&\hspace{-1cm}\times\hspace{-1mm}\int[\mathscr{D}x\mathscr{D}y]
\exp{\Bigl\{\frac{i}{\hslash}\hspace{-1mm}\int\limits_{t_0}^{t_1}\bigl(\tfrac{1}{2}\dot{x}^2+\tfrac{1}{2}\dot{y}^2
-\tfrac{1}{2}y^2-xp_{ycl}\bigr)dt\Bigr\}}.
\end{eqnarray}
The path integral in the configuration space we have constructed
is quadratic and hence can be computed explicitly. The normalized
transition amplitude is
$$
\mathbf{A}_D(x_1,y_1,t_1|x_0,y_0,t_0)=\frac{\mbox{\large{e}}^{\frac{i}{\hslash}\mathbf{S}_{Dcl}}}{2\pi i\hslash}\,\left|
\begin{array}{cc}
\dfrac{\partial^2 \mathbf{S}_{Dcl}}{\partial x_1\,\partial x_0} & \dfrac{\partial^2 \mathbf{S}_{Dcl}}{\partial x_1\,\partial y_0} \\
\vspace{-2mm}\\
\dfrac{\partial^2 \mathbf{S}_{Dcl}}{\partial y_1\,\partial x_0} & \dfrac{\partial^2 \mathbf{S}_{Dcl}}{\partial y_1\,\partial y_0}
\end{array}
\right|^{\frac 12}\hspace{-2mm},
$$
where
\begin{eqnarray*}
\mathbf{S}_{Dcl}&=&
\frac{(x_1-x_0)^2}{2(t_1-t_0)}+\tan^2{\left(\frac{t_1-t_0}{2}\right)}\,\frac{(y_1+y_0)^2}{2(t_1-t_0)}\\
&+&\frac{\tan{\left(\frac{t_1-t_0}{2}\right)}}{(t_1-t_0)}\,(x_1-x_0)(y_1+y_0)-(x_1y_1-x_0y_0)\\
&+&\frac{3}{4\sin{(t_1-t_0)}}\bigl[(y_1^2+y_0^2)\cos{(t_1-t_0)}-2y_1y_0\bigr]\\
&-&\frac{(t_1-t_0)}{4\sin^2{(t_1-t_0)}}\bigl[(y_1^2+y_0^2)-2y_1y_0\cos{(t_1-t_0)}\bigr]\,.
\end{eqnarray*}

\section{Conclusion, discussion and outlook}\label{5}

In the paper we have developed a new quantization method that
generalizes the conventional path integral approach. Throughout
the paper we considered only the nonrelativistic quantum mechanics
of spinless systems. However, the generalization to the field
theory is rather straightforward. We have just to pass from the
space of particle positions to the space of field configurations.

The mathematical language of our exposition respects the Vladimir
Arno\v{l}d ``principle of minimal generality.'' Formulas are
mostly written down in one (local) chart. From this, however, one
can ascend to a global, coordinate free description employing
bundles and jet prolongations. In order that we did not cloud up
the main idea of the paper, we did not follow such fluffy
approach. However, in the future it could be useful to analyze
obstructions to the implementation of our method which arise from
the global properties of the underlying space-time geometry.

Special attention was paid to the stringy quantization of
dissipative systems and of the Douglas system. The latter cannot
be quantized by any known quantization technique while the former
can. In fact, dissipative (friction force) systems were studied
extensively in the past and several approaches to their
quantization were proposed. The stringy quantization was compared
with three of them, Caldirola-Kanai, Bateman and Kostin, and it
was shown that it either gives better results (in the first two
cases) or is simpler to apply (in the third case).

\subsection{Stringy \emph{versus} Caldeira-Leggett model}

As mentioned in the introduction, the quantization proposed here
shares a common property with the well established approaches of
Caldirola-Kanai, Bateman, Kostin and others: they all fail to
cover decoherence phenomena. On the other hand, as we have seen,
the stringy quantization produces a quantum propagator that does
not respect the memoryless Chapman-Kolmogorov equation. We believe
that this essential feature of the theory is measuring/reflecting
the non-Lagrangeanity of the system on the quantum level.

From the author's point of view one can rise a conceptual
objection also against the particle-plus-environment model by
Caldeira and Leggett \cite{particle+environment}. Their approach
is microscopical except the way in which the spectral density
$\rho_D(\omega)$ for the reservoir degrees of freedom is
introduced. This density is not obtained from any kind of
microscopical theory and the form it assumes is motivated solely
by the necessity to obtain the dissipative term $-\kappa\dot{q}$
in the effective theory. On the top of it, the interaction
Hamiltonian is such that it produces, after integrating out the
reservoir degrees of freedom, an additional Langevin stochastic
force $F(t)$. Thus, the effective classical motion of the particle
is governed by the Brownian equation of motion $\dot{p}=-\kappa
p+F(t)$, which is conceptually different from the physical
situation considered in section \ref{4.1}. It is also worth to
point out that the Caldeira-Leggett model cannot describe
satisfactorily the friction force proportional to $-p^A$ with
$A\neq 1$. There is no doubt that for $A=1$ the model gives better
results than all effective theories mentioned before including the
stringy one, but it is fair to say that it describes a different
physical phenomenon, namely the quantum Brownian motion of a
particle which is in thermal equilibrium with a reservoir.

\subsection{Surface functional integral \emph{versus} Feynman-Vernon}

The goal of our paper was to write down a functional integral
formula for the quantum transition amplitude in terms of the
underlying classical equation of motion. However, we did not
eliminate the artificial notions of pure states and classical
action for non-Lagrangian systems. In the surface functional
integral (\ref{FeynmanIII}), these notions were implicitly
present.

A possible alternative to our approach consists in expressing the
transition probability (not the probability amplitude!) using the
functional integral. The probability that the system evolves from
the mixed state $\rho_{_0}(q_0,q_0^\prime,t_0)$ at the time $t_0$
to the mixed state $\rho_{_1}(q_1,q_1^\prime,t_1)$ at some later
time $t_1$ is
\begin{eqnarray*}
\mathbf{P}_{\rho_{_0}\rightarrow\rho_{_1}} &\propto& \int dq_0dq_0^\prime dq_1dq_1^\prime\rho_{_0}(q_0,q_0^\prime,t_0)
\rho_{_1}(q_1,q_1^\prime,t_1)\times\\
&\times& \int[\mathscr{D}\tilde{\gamma}][\mathscr{D}\tilde{\gamma}^\prime]\exp\Bigl\{\frac{i}{\hslash}
\Bigl(\int\limits_{\tilde{\gamma}}-\int\limits_{\tilde{\gamma}^\prime}\Bigr)(pdq-Hdt)\Bigr\}\,.
\end{eqnarray*}
Here $\tilde{\gamma}$ and $\tilde{\gamma}^\prime$ are curves in
the extended phase space whose $q$-projections connect $q_0$ with
$q_1$ and $q_0^\prime$ with $q_1^\prime$ respectively. Using the
Stokes theorem one is able to convert the difference of the line
integrals of the one-form $pdq-Hdt$ into the surface integral of
the two-form $\Omega=d(pdq-Hdt)$. Explicitly,
\begin{equation}\label{loop3}
\Bigl(\int\limits_{\tilde{\gamma}}-\int\limits_{\tilde{\gamma}^\prime}\Bigr)pdq-Hdt=-\int\limits_\Sigma\Omega-
\Bigl(\int\limits_{\lambda_1}-\int\limits_{\lambda_0}\Bigr) pdq\,,
\end{equation}
where $\Sigma$ represents again a map from the parametric space
$(t,s)\in[t_0,t_1]\times[0,1]$ to the extended phase space,
$$
\Sigma: (t,s)\mapsto
\Sigma(t,s)\equiv\bigl(q_\Sigma(t,s),p_\Sigma(t,s),t_\Sigma(t,s)=t\bigr)\,,
$$
such that
$\partial\Sigma=\tilde{\gamma}^\prime(t)-\lambda_1(s)-\tilde{\gamma}(t)+\lambda_0(s)$.
The sideways $\lambda$-boundary curves\footnote{The situation here
strongly resembles that discussed in section \ref{3}. However,
there is one substantial difference between the two cases. In the
present case, we allow the $\lambda$-curves to vary in both
coordinates $q$ and $p$, only the time $t$ must stay unchanged
along them. In the case considered in section \ref{3} we were more
stringent: the $\lambda$-curves were allowed to vary only with
respect to $p$ and had to stay unchanged with respect to both $q$
and $t$.} of $\Sigma$
\begin{equation*}
\lambda_0(s)\equiv\Sigma(t_0,s)\ \ \ \ \mbox{and}\ \ \ \
\lambda_1(s)\equiv\Sigma(t_1,s)
\end{equation*}
live in the instant phase spaces of
$\mathbb{R}^{2n+1}[q^a,p_a,t]$, i.e. in the submanifolds
$\mathbb{R}^{2n}[q^a,p_a,t_0]$ and $\mathbb{R}^{2n}[q^a,p_a,t_1]$.
Moreover, as the intrinsic parameter $s$ varies from $0$ to $1$,
the $q_\Sigma$-components of $\lambda_0(s)$ and $\lambda_1(s)$
vary from $q_0$ to $q_0^\prime$ and from $q_1$ to $q_1^\prime$
respectively.

Let us denote by $\mathscr{U}$ the space of all $\Sigma$-maps for which
$$
{q_\Sigma(t_0,0)=q_0,  \atop q_\Sigma(t_0,1)=q_0^\prime,}\ \
{q_\Sigma(t_1,0)=q_1,  \atop q_\Sigma(t_1,1)=q_1^\prime,}\ \
{\lambda_0(s)\subset\mathbb{R}^{2n}[q,p,t_0], \atop
\lambda_1(s)\subset\mathbb{R}^{2n}[q,p,t_1].}
$$
Since the right hand side of (\ref{loop3}) depends only on
$\Sigma$ and its two boundaries, the double path integral
$\int[\mathscr{D}\tilde{\gamma}][\mathscr{D}\tilde{\gamma}^\prime]\cdots$
entering $\mathbf{P}_{\rho_{_0}\rightarrow\rho_{_1}}$ can be
rewritten as a surface functional integral
$\int[\mathscr{D}\Sigma]\cdots$,
\begin{eqnarray}\label{Feynman-Vernon}
\vspace{5mm}
\mathbf{P}_{\rho_{_0}\rightarrow\rho_{_1}}\hspace{-2mm}
&\propto&\hspace{-2mm} \int\hspace{-1mm} dq_0dq_0^\prime
dq_1dq_1^\prime\rho_{_0}
(q_0,q_0^\prime,t_0)\rho_{_1}(q_1,q_1^\prime,t_1)\times\nonumber\\
\hspace{-2mm}&\times&\hspace{-2mm}
\int\limits_{\mathscr{U}}[\mathscr{D}\Sigma]\exp\Bigl\{-\frac{i}{\hslash}\Bigl[
\int\limits_\Sigma\Omega+
\Bigl(\int\limits_{\lambda_1}\hspace{-1mm}-\hspace{-1mm}\int\limits_{\lambda_0}\Bigr)
pdq\Bigr]\Bigr\}\,.
\end{eqnarray}
The formula we have just arrived at was derived under the
assumption that $\Omega=d(pdq-Hdt)$. However, it is clear that the
expression in the exponent is free of any reference to the
Hamiltonian and requires just the classical equations of motion.
Therefore it seems reasonable to postulate the probability formula
(\ref{Feynman-Vernon}) also for non-Lagrangian systems.

The approach we shortly presented here does not use the notion of
pure states of a non-Lagrangian system. It resembles the
Feynman-Vernon approach \cite{feynman-vernon}, in which one
introduces the influence functional in the presence of dissipative
forces. Our probability formula (\ref{Feynman-Vernon}) uses the
surface functional integral in the extended phase space, while in
the Feynman-Vernon approach one just has to calculate a double
path integral in the configuration space. There is a chance that
after a certain discretization we will be able to convert the
surface functional integral into the double path integral in the
configuration space (see appendix \ref{B}), and as a result, we
will find the explicit form of the influence functional. Work on
this topic is in progress.

\begin{acknowledgements}
This research was supported by M\v S SR CERN-Fellowship Program
and VEGA Grant 1/1008/09. Special thanks go to Vladim\'ir Balek
for his interest and fruitful discussions
we have had in the course of the work.\\
\\

\centerline{$\mathscr{A.}$\ \ \ \ \ \ $\mathscr{M.}$\ \ \ \ \ \
$\mathscr{D.}$\ \ \ \ \ \ $\mathscr{G.}$}
\end{acknowledgements}

\appendix

\section{Stringy variational principle}\label{A}

In section \ref{3} we shortly outlined how the transition
amplitude (\ref{FeynmanIII}) can be related to the stringy
variational principle, with the action $\mathscr{S}$ defined in
(\ref{action}). In what follows we will perform a complete
variation of $\mathscr{S}$, with the domain extended from
$\mathscr{U}_{\gamma_{cl}}$ to a wider class
$\mathscr{U}:=\bigcup_{\tilde{\gamma}_0}\bigcup_{\tilde{\gamma}_1}
\mathscr{U}_{(\tilde{\gamma}_1,\tilde{\gamma}_0)}$. The stringy
set $\mathscr{U}$ contains all surfaces in the extended phase
space trapped between the submanifolds  with fixed $(q_0,t_0)$ and
$(q_1,t_1)$ (including separate histories, regarded as shrunk
surfaces). In the course of variation the classical dynamics
(\ref{eq.}) we have started from will be recovered.

\begin{figure}
\begin{center}
\includegraphics[height=11.18cm,width=8.5cm]{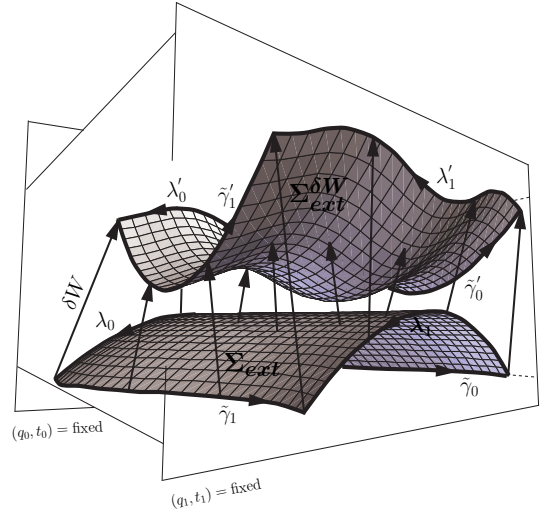}
\vspace{-3cm} \caption{Vector field $W$ acting infinitesimally in
the extended phase space. As a result, the initial surface
$\Sigma_{ext}\in\mathscr{U}_{(\tilde{\gamma}_1,\tilde{\gamma}_0)}$
moves to some nearby surface $\Sigma_{ext}^{\delta
W}\in\mathscr{U}_{(\tilde{\gamma}_1^{_{'}},\tilde{\gamma}_0^{_{'}})}$,
and its boundary $\partial\Sigma_{ext}=\tilde{\gamma}_1-\lambda_1-
\tilde{\gamma}_0+\lambda_0$ is transported to a new boundary
$\partial\Sigma_{ext}^{\delta
W}=\tilde{\gamma}_1^{_{'}}-\lambda_1^{_{'}}-
\tilde{\gamma}_0^{_{'}}+\lambda_0^{_{'}}$. Since the variational
vector field is assumed to preserve $\mathscr{U}$, $W$ stays
tangential to the momentum submanifolds with fixed $(q_0,t_0)$ and
$(q_1,t_1)$.} \label{flow}
\end{center}
\end{figure}

Suppose we have an extremal surface
$\Sigma_{ext}\in\mathscr{U}_{(\tilde{\gamma}_1,\tilde{\gamma}_0)}\subset\mathscr{U}$
and a variational vector field $W$ defined in its neighborhood
such that the flow of $W$ preserves $\mathscr{U}$. By definition,
$W$ moves
$\Sigma_{ext}\in\mathscr{U}_{(\tilde{\gamma}_1,\tilde{\gamma}_0)}$
to some other worldsheet $\Sigma_{ext}^{\delta
W}\in\mathscr{U}_{(\tilde{\gamma}_1^{_{'}},\tilde{\gamma}_0^{_{'}})}$,
where $\delta$ is an infinitesimal increment of the parameter of
the flow generated by $W$. The situation is depicted schematically
in Figure \ref{flow}.

The extremality of $\Sigma_{ext}$ means that for all variational
$W$-fields it holds
$$
0=\lim\limits_{\delta\rightarrow 0}\frac{\mathscr{S}\bigl(\Sigma_{ext}^{\delta W}\bigr)-\mathscr{S}\bigl
(\Sigma_{ext}\bigr)}{\delta}=\hspace{-1.8mm}\int\limits_{\Sigma_{ext}}\hspace{-1.8mm}\mathcal{L}_{W}\Omega\,,
$$
where $\mathcal{L}_W$ stands for the Lie derivative. Let us
compute the surface integral on the left hand side:
\begin{eqnarray}
&&\hspace{-2cm} \label{variation}
\int\limits_{\Sigma_{ext}}\hspace{-1.8mm}\mathcal{L}_{W}\Omega=\hspace{-1.8mm}\int\limits_{\Sigma_{ext}}
\hspace{-1.8mm}d\bigl(W\lrcorner\,\Omega\bigr)+W\lrcorner\,d\Omega=\nonumber\\
&&\hspace{1cm}=\int\limits_{\partial\Sigma_{ext}}\hspace{-2.5mm}W\lrcorner\,\Omega+\hspace{-1.8mm}\int
\limits_{\Sigma_{ext}}\hspace{-1.8mm}W\lrcorner\,d\Omega\,,
\end{eqnarray}
where the the symbol $\lrcorner$ denotes the inner product
(contraction) of a vector with a differential form. The first term
on the right hand side depends only the values of $\Omega$ on the
boundary of $\Sigma_{ext}$ and can be recast into the form
$$
\int\limits_{\partial\Sigma_{ext}}\hspace{-2.5mm}W\lrcorner\,\Omega=\Bigl(\int\limits_{\tilde{\gamma}_1}-
\int\limits_{\tilde{\gamma}_0}+\int\limits_{\lambda_0}-\int\limits_{\lambda_1}\Bigr)\,W\lrcorner\,\Omega\,.
$$
The last two terms here give individually zero contributions, as
can be seen from the following quick consideration: the vector
field $W$ is assumed to preserve $\mathscr{U}$ and therefore its
restrictions to the $\lambda$-boundaries are
$$
W\hspace{-0.35mm}\bigr|_{\lambda_{0}}\hspace{-1.9mm}=W_a\bigl(p,q_0,t_0\bigr)\frac{\partial}{\ \partial_{p_a}}\,,\ \ \
W\hspace{-0.35mm}\bigr|_{\lambda_{1}}\hspace{-1.9mm}=W_a\bigl(p,q_1,t_1\bigr)\frac{\partial}{\ \partial_{p_a}}.
$$
The one-form we are integrating is
$W\lrcorner\,\Omega=W_a\bigl(dq^a-(p^a/m)dt\bigr)$, but since both
$q$ and $t$ stay unchanged on $\lambda$, the integral is zero. As
a result, to annihilate the boundary term for all variational
fields $W$ we are forced to chose the curves
$\tilde{\gamma}_0=(\tilde{q}_0(t),\tilde{p}_0(t),t)$ and
$\tilde{\gamma}_1=(\tilde{q}_1(t),\tilde{p}_1(t),t)$ in the
extended phase space in such a way that their instant tangent
vectors annihilate $\Omega$ at any moment $t\in[t_0,t_1]$. Thus,
it must hold
\begin{subequations}\label{boundary}
\begin{equation}
0=\frac{d\tilde{\gamma}}{dt}\lrcorner\,\Omega\,,
\end{equation}
or, equivalently,
\begin{eqnarray}
&&\hspace{-1cm}
0=(F_a-\dot{p}_a)dq^a+\left(\dot{q}^a-\frac{p^a}{m}\right)dp_a+\nonumber\\
&&\hspace{1.7cm}
+\left(\dot{q}^aF_a-\dot{p}_a\frac{p_a}{m}\right)dt\,,
\end{eqnarray}
\end{subequations}
for both ${\tilde{\gamma}_0}$ and ${\tilde{\gamma}_1}$. Here one
recognizes the dynamical equations (\ref{eq.}) as desired.

From the boundary term analysis and the assumption about the
uniqueness of the history $\gamma_{cl}$ which we adopted from the
very beginning we can conclude that
$\Sigma_{ext}\in\mathscr{U}_{(\gamma_{cl},\gamma_{cl})}\subset\mathscr{U}_{\gamma_{cl}}$.
Hence $\Sigma_{ext}$ has the topology of a closed string attached
to the classical trajectory $\gamma_{cl}$. To specify its shape
the second term in (\ref{variation}) must be employed. The
resulting variational equation, written in a coordinate-free
notation, is
\begin{subequations}\label{bulk}
\begin{equation}
0=d\Omega\,\bigl(\partial_t\,\Sigma_{ext},\,\partial_s\,\Sigma_{ext},\,.\,\bigr)\,.
\end{equation}
This is equivalent to the following system of partial differential
equations for the unknown functions $q^a(t,s)$ and $p_a(t,s)$
(indices $a$ and $k$ run from $1,\dots, n$):
\begin{equation}
\begin{aligned}
0&= \frac{\partial q^{k}}{\partial s}\,\frac{\partial F_k}{\partial p_{a}}\,,\\
0&= \frac{\partial q^{k}}{\partial s}\,\Bigl(\frac{\partial F_{a}}{\partial q^{k}}-\frac{\partial F_{k}}{\partial q^{a}}\Bigr)
+\frac{\partial p_{k}}{\partial s}\,\frac{\partial F_{a}}{\partial p_{k}}\,.
\end{aligned}
\end{equation}
\end{subequations}
One solution of these equations, satisfying all boundary
conditions, is trivial. It is the shrunk surface
$\Sigma_{ext}(t,s)=\gamma_{cl}(t)$. After recalling the assumption
of uniqueness of $\gamma_{cl}$ once more, we can see that this is
the {\it only} solution of (\ref{bulk}). If there existed a closed
unshrunk extremal surface $\Sigma_{ext}$, the initial set of
equations of motion (\ref{eq.}) would have at least one
one-parametrical family of classical solutions between the given
pair of endpoints.

The variational principle we described above operates on a wider
stringy class $\mathscr{U}$ than it is in fact necessary. The
surface integral formula (\ref{FeynmanIII}) requires just the
restricted subset $\mathscr{U}_{\gamma_{cl}}\subset\mathscr{U}$.
The transition to $\mathscr{U}_{\gamma_{cl}}$ is advisable for two
reasons. First, in the Lagrangian case with $\Omega=d(pdq-Hdt)$ we
obtain an equivalence (modulo additive constant) between the least
action principle using $\mathscr{S}$ and the standard Hamilton
least action principle. Explicitly,
$$
\Sigma\in\mathscr{U}_{(\tilde{\gamma},\gamma_{cl})}\subset\mathscr{U}_{\gamma_{cl}}\mapsto\mathscr{S}(\Sigma)=
\Bigl(\,\int\limits_{\tilde{\gamma}}-\int\limits_{\gamma_{cl}}\,\Bigr)(pdq-Hdt)\,.
$$
Second, the wider class $\mathscr{U}$ contains plenty of
degenerate (shrunk) surfaces, the histories $\tilde{\gamma}$.
These are obviously stationary surfaces of $\mathscr{S}$ when
being varied within $\mathscr{U}$, since
$$
\mathscr{S}(\tilde{\gamma}^{\delta W})-\mathscr{S}(\tilde{\gamma})=\hspace{-1mm}\int\limits_{\tilde{\gamma}^{\delta W}}
\hspace{-1mm}\Omega\,-\int\limits_{\tilde{\gamma}}\hspace{-0mm}\Omega=0-0=0
$$
for any variational vector field $W$. However, only one of these
histories, namely $\gamma_{cl}$, satisfies both equations
(\ref{boundary}) and (\ref{bulk}) at the same time. This
fictitious problem is avoided when one works from the beginning
with the stringy subclass
$\mathscr{U}_{\gamma_{cl}}\subset\mathscr{U}$.

\section{Surface functional integral -- computational details}\label{B}

\begin{figure}
\begin{center}
\includegraphics[height=11.18cm,width=8.5cm]{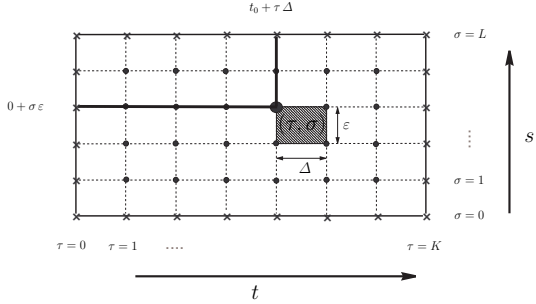}
\vspace{-6cm}
\caption{Rectangular nodal web in the parametric space
$[t_0,t_1]\times[0,1]$ of a surface
$\Sigma\in\mathscr{U}_{(\tilde{\gamma},\gamma_{cl})}$. Each
elementary tile encloses the area
$\varDelta\cdot\varepsilon=(t_1-t_0)/(KL)$ (at the end, the
numbers $K$ and $L$ will be sent to infinity). The points marked
by crosses are constrained by the conditions (\ref{constraints}).}
\label{grid}
\end{center}
\end{figure}

Let us explain here in some detail how the surface functional
integral is computed. To be as tangible as possible consider a
one-dimensional system only, so that the extended phase space will
be the three dimensional space $\mathbb{R}^3[q,p,t]$ (more
dimensions represent only a technical problem). The particle is
supposed to move under the combined action of the potential $U(q)$
and the friction force with a general $A$-power law. The dynamical
equations are
$$
\dot{q}=p\,,\ \ \dot{p}=-\frac{d}{dq}U(q)-\kappa\,\dot{p}^{A}\,,
$$
and in addition to them, we require that the particle satisfies
the boundary conditions $q(t_0)=q_0$ and $q(t_1)=q_1$. According
to the definition (\ref{Omega}), the two-form $\Omega$ is
\begin{equation}\label{alpha}
\Omega=d\bigl(pdq-\frac{1}{2}p^2dt-U(q)dt\bigr) -\kappa\, p^{A}dq\wedge dt\,.
\end{equation}
Our aim is to compute the surface functional integral
(\ref{FeynmanIII}) over the stringy set
$\mathscr{U}_{\gamma_{cl}}=\bigcup_{\tilde{\gamma}}\mathscr{U}_{(\tilde{\gamma},\gamma_{cl})}$.
The direct application of (\ref{surface}) and (\ref{alpha})
together with the Stokes theorem yields
\begin{eqnarray}
& & \hspace{-5mm}\int\limits_{\mathscr{U}_{\gamma_{cl}}}\hspace{-1mm}[\mathscr{D}\Sigma]\exp{\Bigl\{\frac{i}{\hslash}
\int\limits_{\Sigma}\Omega\Bigr\}}=\label{giuseppe}\\
& & \hspace{-2mm}=\int[\mathscr{D}\tilde{\gamma}]\exp{\Bigl\{\frac{i}{\hslash}\Bigl(\,\int\limits_{\tilde{\gamma}}-
\int\limits_{\gamma_{cl}}\,\Bigr)\bigl(pdq-\frac{1}{2}p^2dt-U(q)dt\bigr)\Bigr\}}\times\nonumber\\
& & \hspace{-2mm}\times\hspace{-4mm}\int\limits_{\ \ \ \mathscr{U}_{(\tilde{\gamma},\gamma_{cl})}}\hspace{-5mm}
[\mathscr{D}\Sigma]\exp{\Bigl\{-\frac{i}{\hslash}\kappa\int\limits_{\Sigma}p^{A}dq\wedge dt\Bigr\}}\,.\nonumber
\end{eqnarray}
The nontrivial part of this is the functional integral over the
stringy subset $\mathscr{U}_{(\tilde{\gamma},\gamma_{cl})}$. As
mentioned earlier,
$\Sigma\in\mathscr{U}_{(\tilde{\gamma},\gamma_{cl})}$ is a map
$$
\Sigma:[t_0,t_1]\times[0,1]\rightarrow\mathbb{R}^3[q,p,t]\,,\ \ \ (t,s)\mapsto\Sigma(t,s)\,,
$$
such that for $\forall t\in[t_0,t_1]$ and $\forall s\in[0,1]$
there holds
\begin{equation}\label{constraints}
{\Sigma(t,0)=\gamma_{cl}(t) \atop \Sigma(t,1)=\tilde{\gamma}(t)\
\,}\ \ \ \ \mbox{and}\ \ \ \
{\Sigma(t_0,s)\in\mathbb{R}^3[q_0,p,t_0] \atop
\Sigma(t_1,s)\in\mathbb{R}^3[q_1,p,t_1]}\,.
\end{equation}
To proceed further with the functional integral in question,
introduce a set of regularly distributed nodal points in the
parameter space,
$$
\bigl\{(t_0+\tau\,\varDelta,\sigma\,\varepsilon)\in\bigr[t_0,t_1]\times[0,1]\}\,.
$$
The points of the set are labeled by two discrete indices, the
time index $\tau=0,\dots, K$ and the space index $\sigma=0,\dots
L$, see Figure \ref{grid}. In this way we obtain for any
$\Sigma$-map from $\mathscr{U}_{(\tilde{\gamma},\gamma_{cl})}$ a
discretized $(K+1)\times (L+1)$-tuple
$$
\Sigma\leftrightsquigarrow\bigl\{\Sigma(t_0+\tau\varDelta,\sigma\varepsilon)\equiv\bigl(q_{(\tau,\sigma)},
p_{(\tau,\sigma)},
t_0+\tau\varDelta\bigr)\bigr\}_{(\tau,\sigma)=(0,0)}^{(K,L)}\,.
$$
The boundary values are required to be consistent with
(\ref{constraints}). Explicitly, for all indices $\tau$ and
$\sigma$ it must hold
$$
{q_{(\tau,0)}=q_{cl}(t_0+\tau\varDelta), \atop
p_{(\tau,0)}=p_{cl}(t_0+\tau\varDelta),}\ \ \
{q_{(\tau,L)}=\tilde{q}(t_0+\tau\varDelta), \atop
p_{(\tau,L)}=\tilde{p}(t_0+\tau\varDelta),}\ \ \
{q_{(0,\sigma)}=q_0, \atop q_{(K,\sigma)}=q_1.}
$$
The discretization of $\Sigma$ enables us to approximate the
integral of $p^{A}dq\wedge dt$ as follows \footnote{From now on we
are using a new symbol $\fint$ for all discretized integrals. In
the continuum limit $K$ and $L$ approach independently infinity
and $\fint\to\int$. However, we must keep in mind that we must
perform simultaneously the limits $\varDelta \to 0$ and
$\varepsilon \to 0$ so that the quantities $K\varDelta=t_1-t_0$
and $L\varepsilon=1$ stay finite.}:
\begin{eqnarray}
\fint\limits_\Sigma& &\hspace{-6mm} p^{A}dq\wedge dt:=\sum\limits_{\tau=0}^{K-1}\sum\limits_{\sigma=0}^{L-1}\varDelta\varepsilon\
\Bigl\{p^A_{(\tau,\sigma)} \frac{q_{(\tau,\sigma+1)}-q_{(\tau,\sigma)}}{\varepsilon}\Bigr\}=\nonumber\\
&=&\sum\limits_{\tau=1}^{K-1}\sum\limits_{\sigma=1}^{L-1}\,\varDelta\ \bigl\{q_{(\tau,\sigma)}\bigl(\,p^A_{(\tau,\sigma-1)}-
p^A_{(\tau,\sigma)}\,\bigr)\bigr\}+\nonumber\\
&+&\sum\limits_{\tau=1}^{K-1}\,\varDelta\bigl\{q_{(\tau,L)}\,p^A_{(\tau,L-1)}-q_{(\tau,0)}\,p^A_{(\tau,0)}\bigr\}\label{discr.1}\,.
\end{eqnarray}

Formally, the discretized functional integral over all stringy
configurations from $\mathscr{U}_{(\tilde{\gamma},\gamma_{cl})}$
is a multiple integral over all unconstrained variables
$x_{(\tau,\sigma)}, p_{(\tau,\sigma)}$ which are needed to specify
$\Sigma$. The only problematic part, as usual, is the choice of an
appropriate integration measure. For the reason that will become
clear in a moment, we choose the measure as
\begin{equation}\label{discr.2}
\hspace{-3mm}\fint\limits_{\ \ \
\mathscr{U}_{(\tilde{\gamma},\gamma_{cl})}}\hspace{-5mm}[\mathscr{D}\Sigma]:=\hspace{-2mm}
\int\limits_{-\infty}^{+\infty}\
\prod\limits_{\tau=1}^{K-1}\prod\limits_{\sigma=1}^{L-1}\,\Bigl(\frac{A\kappa\varDelta}{2\pi
i\hslash}\,p^{A-1}_{(\tau,\sigma)}\Bigr)\,
dq_{(\tau,\sigma)}\,dp_{(\tau,\sigma)}.\hspace{-5mm}
\end{equation}
The first step, when dealing with the discretized functional
integral
$$
\mathrm{Int}\equiv\hspace{-4mm}\fint\limits_{\ \ \ \mathscr{U}_{(\tilde{\gamma},\gamma_{cl})}}\hspace{-5mm}[\mathscr{D}\Sigma]
\exp{\Bigl\{-\frac{i}{\hslash}\kappa\fint\limits_{\Sigma}p^{A}dq\wedge dt\Bigr\}}\,,
$$
consists in the integration over the internal stringy positions
$dq_{(\tau,\sigma)}$. After this integration, the integrand
transforms into a chain of delta functions,
\begin{eqnarray*}
\mathrm{Int}\hspace{-0.5mm}&=&\hspace{-2mm}\int\limits_{-\infty}^{+\infty}\prod\limits_{\tau=1}^{K-1}\prod\limits_{\sigma=1}^{L-1}\,
\bigl(A\,p^{A-1}_{(\tau,\sigma)}\bigr)\,dp_{(\tau,\sigma)}\,\mbox{\large{$\boldsymbol{\delta}$}}\bigl(p^A_{(\tau,\sigma-1)}-
p^A_{(\tau,\sigma)}\bigr)\times\\
&\times& \exp{\Bigl\{-\frac{i}{\hslash}\kappa
\sum\limits_{\tau=1}^{K-1}\,\varDelta\bigl[q_{(\tau,L)}\,p^A_{(\tau,L-1)}-q_{(\tau,0)}\,
p^A_{(\tau,0)}\bigr]\Bigr\}}\,.
\end{eqnarray*}
The next step is the integration over the internal stringy momenta
$dp_{(\tau,\sigma)}$. After performing this trivial integration we
arrive at the expression
$$
\mathrm{Int}=\exp{\Bigl\{-\frac{i}{\hslash}\kappa
\sum\limits_{\tau=1}^{K-1}\,\varDelta\bigl[q_{(\tau,L)}\,p^A_{(\tau,0)}-q_{(\tau,0)}\,
p^A_{(\tau,0)}\bigr]\Bigr\}}\,.
$$
In the exponent there appears the discretized version of the
integral
$$
\int\limits_{t_0}^{t_1}\,dt\bigl\{\tilde{q}(t)\,p^A_{cl}(t)-q_{cl}(t)\,p^A_{cl}(t)\bigr\}\,,
$$
where $\tilde{q}(t)$ stands for the $q$-projection of
$\tilde{\gamma}(t)$. Consequently, after returning back to the
continuum limit we obtain the following important result:
\begin{eqnarray}\label{int.}
&&\hspace{-1.2cm}\int\limits_{\ \ \
\mathscr{U}_{(\tilde{\gamma},\gamma_{cl})}}\hspace{-5mm}[\mathscr{D}\Sigma]
\exp{\Bigl\{-\frac{i}{\hslash}\kappa\int\limits_{\Sigma}p^{A}dq\wedge dt\Bigr\}}=\nonumber\\
&&\hspace{-5mm}=\exp{\Bigl\{-\frac{i}{\hslash}\kappa
\int\limits_{t_0}^{t_1}\,dt\bigl[\tilde{q}(t)\,p^A_{cl}(t)-q_{cl}(t)\,p^A_{cl}(t)
\bigr]\Bigr\}}\,.
\end{eqnarray}
After substituting (\ref{int.}) into the initial formula
(\ref{giuseppe}) we recover a path integral in the extended phase
space. The integral is quadratic in momenta, with the discretized
(standard) Liouville measure
$$
\fint[\mathscr{D}\tilde{\gamma}]\ \cdots
=\int\limits_{-\infty}^{+\infty}\frac{d\tilde{p}_K}{2\pi\hslash}\,\prod\limits_{\tau=1}^{K-1}
\frac{d\tilde{p}_{\tau}\,d\tilde{q}_{\tau}}{2\pi\hslash}\
\cdots\,,
$$
The integration over momenta can be carried out explicitly and we
finally obtain a path integral in the configuration space only
(tildes are removed from the position variables),
\begin{eqnarray}
&&\hspace{-8mm}\int\limits_{\mathscr{U}_{\gamma_{cl}}}[\mathscr{D}\Sigma]
\exp{\Bigl\{\frac{i}{\hslash}\int\limits_{\Sigma}\Omega\Bigr\}}=\nonumber\\
&&\hspace{-5mm}=\exp{\Bigl\{-\frac{i}{\hslash}\int\limits_{t_0}^{t_1}\Bigl(\frac{1}{2}
\dot{q}_{cl}^2-U(q_{cl})-\kappa q_{cl}\,p^A_{cl}\Bigr)dt\Bigr\}}\times\nonumber\\
&&\hspace{-5mm}\times\int[\mathscr{D}q]\exp{\Bigl\{\frac{i}{\hslash}\int\limits_{t_0}^{t_1}
\Bigl(\frac{1}{2}\dot{q}^2-U(q)-\kappa
q\,p^A_{cl}\Bigr)dt\Bigr\}}\,.
\end{eqnarray}
This formula served us as a guide when introducing the classical
action (\ref{A-action}).

Let us make one final comment. In our functional measure
(\ref{discr.2}) no integration over the momenta $p_{(0,\sigma)}$
and $p_{(K,\sigma)}$ was prescribed. These momenta are however
needed when the boundary of the discretized surface
$\Sigma\in\mathscr{U}_{\gamma_{cl}}$ is specified. They define the
auxiliary curves $\lambda_0(s)$ and $\lambda_1(s)$ introduced in
section \ref{3}. The curves were chosen completely arbitrarily,
but as seen from (\ref{discr.1}), the result is not affected by
them at all. Hence, since everything substantial is independent of
$p_{(0,\sigma)}$ and $p_{(K,\sigma)}$, it is justifiable to
discard these quantities from the functional measure
$[\mathscr{D}\Sigma]$. If we would not do that for some reason,
they will integrate into an artificial infinite factor, which will
be removed anyway after we apply the normalization conditions
(\ref{normalization}).

\end{document}